\documentclass[aps,prl,superscriptaddress, twocolumn]{revtex4-2}
\usepackage{amsmath}
\usepackage{amssymb}
\usepackage{bm}
\usepackage{color}
\usepackage{appendix}
\usepackage{graphicx}
\usepackage{braket}
\usepackage{listings}
\usepackage{physics}
\usepackage{titlesec}

\titleformat{\part}
  {\normalfont\fontsize{12}{15}\bfseries}{\thepart}{1em}{}

\begin{document}
\title{Suppression of Pulsed Dynamic Nuclear Polarization by Many-Body Spin Dynamics}
\author{Kento Sasaki}
\email{kento.sasaki@phys.s.u-tokyo.ac.jp}
\affiliation{Department of Physics, The University of Tokyo, Bunkyo-ku, Tokyo, 113-0033, Japan}
\author{Eisuke Abe}
\email{eisuke.abe@riken.jp}
\affiliation{RIKEN Center for Quantum Computing, Wako, Saitama 351-0198, Japan}
\date{\today}
\begin{abstract}
We study a mechanism by which nuclear hyperpolarization due to the polarization transfer from a microwave-pulse-controlled electron spin is suppressed. 
From analytical and numerical calculations of the unitary dynamics of multiple nuclear spins, we uncover that, combined with the formation of the dark state within a cluster of nuclei,
coherent higher-order nuclear spin dynamics impose limits on the efficiency of the polarization transfer even in the absence of mundane depolarization processes such as nuclear spin diffusion and relaxation.
Furthermore, we show that the influence of the dark state can be partly mitigated by introducing a disentangling operation.
Our analysis is applied to the nuclear polarizations observed in $^{13}$C nuclei coupled with a single nitrogen-vacancy center in diamond [Science $\bm{374}$, 1474 (2021) by J. Randall {\it et al.}].
Our work sheds light on collective engineering of nuclear spins as well as future designs of pulsed dynamic nuclear polarization protocols.
\end{abstract}
\maketitle

In the burgeoning field of quantum technology, nuclear spins in solids and molecules have multiple facets.
On the one hand, they function as long-lived qubits, constituting a building block for quantum computers~\cite{K98,ZAS19,PM21,AWR+22}, quantum simulators~\cite{CRJP13,UTW+18,RBvdG+21}, and quantum memories~\cite{BRA+19,PHB+21,RWR+22,HPB+22}.
On the other hand, a small ensemble of nuclear spins can be detected by utilizing an atom-sized quantum sensor such as a nitrogen-vacancy (NV) center in diamond~\cite{AS18};
nanoscale NMR spectroscopy realized this way will have a transformative impact on chemistry, biology, medicine, and other related disciplines~\cite{ARB+19,CHZ+22,ABB22}.
It has also been demonstrated that nuclear spins themselves can serve as sensitive magnetometers~\cite{SdLSC+22}.

In all applications, a prerequisite is to hyperpolarize nuclei at a level far beyond the thermal equilibrium, ideally to unity.
Although the concept of dynamic nuclear polarization (DNP), in which the spin polarization is transferred from electrons to nuclei,
can be traced back to the prediction and demonstration of the Overhauser effect in 1953~\cite{O53,CS53} and various techniques have been developed since then~\cite{HDSW88,HDW88,R-BL16,HNJB18,TJTG19}, 
an additional challenge here is that qubits/sensors are often operated at low magnetic fields ($B_0 <$ 100~mT) and at room temperatures.
A promising approach is to engineer the electron--nuclear interaction via pulsed microwave control of the electron spin.
For instance, the PulsePol sequence has been introduced as a broadband and robust method for the polarization transfer~\cite{SST+18}
and applied to recent experiments on quantum sensing and quantum simulation with NV centers~\cite{SIA18,SWS+20,RBvdG+21}.
Nonetheless, the near-unity hyperpolarization remains elusive, and it is pivotal to understand the suppression mechanisms. 

In this work, we study higher-order many-body nuclear spin dynamics induced by a pulsed DNP sequence itself,
with particular interest in a system with an optically polarizable electron spin embedded in a sparse nuclear spin bath.
The NV center in diamond is a prime example, but other color centers in semiconductors~\cite{SAB+20,ZCCG20}, molecular qubits~\cite{TNN+14,BLM+20,BDL+22},
rare-earth ions in solids~\cite{RWR+22,ZHA+15}, to name a few, are also relevant platforms.
In a model system with a single $S = \frac{1}{2}$ electron spin interacting with multiple $I = \frac{1}{2}$ nuclear spins,
we perform direct computations of the unitary dynamics during PulsePol and derive analytical expressions of the transition amplitudes of the nuclear spins,
which are more involved to obtain via prevailing perturbative expansion methods based on average Hamiltonian theory~\cite{HW68,H76,HN98,LMV10,SvBE10,TNSM15,NHAV19}.
After numerical simulations on the larger nuclear spins,
we analyze the case of $^{13}$C nuclei coupled with a single NV center in diamond, including a concrete dataset from a spin-based quantum simulator reported in~\cite{RBvdG+21}.

The system Hamiltonian under the secular approximation reads
\begin{equation*}
\hat{H} = \sum_{m_S} |m_S\rangle \langle m_S| \otimes
\sum_{l = 1}^{N_\text{nuc}} (\mathalpha{-} f_\text{n} \mathalpha{+} m_S A_{\parallel}^{(l)}) \hat{I}_{z}^{(l)}
\mathalpha{+} m_S A_{\perp}^{(l)} \hat{I}_{x}^{(l)},
\end{equation*}
where $m_S = \pm \frac{1}{2}$ is the electron spin state, $N_\text{nuc}$ is the number of nuclear spins,
$f_\text{n}$ is the nuclear Larmor frequency, $\hat{I}_{z,x}^{(l)}$ are the operators of the $l$-th nuclear spin, 
and $A_{\parallel}$ ($A_{\perp}$) is the scalar hyperfine parameter parallel (perpendicular) to the external magnetic field $B_0$.
Due to the hyperfine interaction, the nuclear spins precess at
$f_\text{p} = \sqrt{ (f_\text{n} \pm A_{\parallel}/2)^2 + (A_\perp/2)^2}$
around the axis tilted by the angle
$\theta=\arctan(A_\perp/2 f_\text{n})$ from the direction of $B_0$.
The inter-nuclear dipolar interactions, which induce a direct flip-flop between a pair of nuclei, are dropped from $\hat{H}$,
partly because in a sparse nuclear bath this time scale is much longer than the pulse sequence time.
More importantly, we show below that higher-order spin dynamics, driving indirect nuclear flip-flop and flip-flip mediated by the electron spin, play a crucial role in depolarization.

\begin{figure}
\begin{center}
\includegraphics{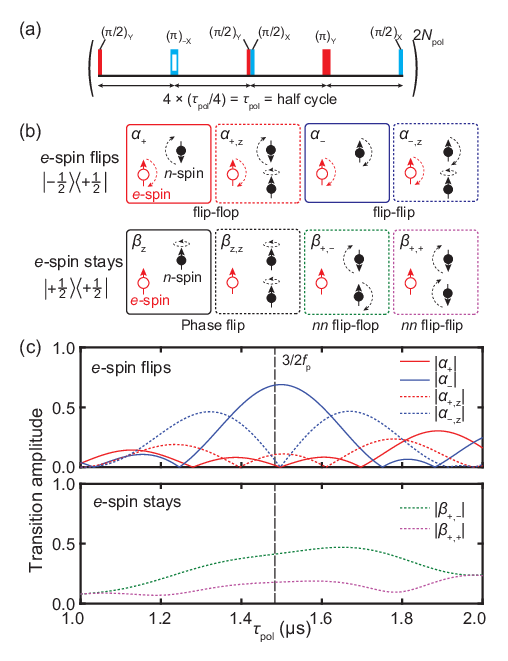}
\caption{
(a) PulsePol sequence.
(b) Possible spin dynamics.
Solid (dotted) frames indicate single (multiple) nuclear spin dynamics.
(c) Spectra of the absolute transition amplitudes.
\label{fig1}
}\end{center}
\end{figure}
The PulsePol sequence [see also Fig.~\ref{fig1}(a)] is given by
\begin{equation*}
\hat{U} = (\hat{U}_x \hat{U}_t \hat{U}_y^2 \hat{U}_t \hat{U}_x \hat{U}_y \hat{U}_t (\hat{U}_x^\dagger)^2 \hat{U}_t \hat{U}_y)^{2N_\text{pol}},
\end{equation*}
where $U_{x,y}=e^{-i \frac{\pi}{2} \hat{S}_{x,y}}$ and $\hat{U}_t = e^{-2\pi i\hat{H} \tau_\text{pol}/4}$ represent
infinitesimally short $\pi/2$ pulses around the $x,y$ axes of the electron spin and the free evolution, respectively.
The electron spin is initialized (optically) into $m_S = +\frac{1}{2}$ in advance and the unit sequence is repeated ${2N_\text{pol}}$ times.
The pulse interval $\tau_\text{pol}$ is chosen so that the polarization transfer will be most efficient at a target frequency $f_\text{t}$,
given by the {\it resonance} condition $f_\text{t} = 3/2\tau_\text{pol}$.

To unveil higher-order spin dynamics in the simplest setting, we start our analysis with $N_\text{nuc}$ = 2 and uniform $A_\perp$.
Furthermore, we assume $A_\parallel$ to be zero, as it primarily contributes to the shift of the precession frequency and does not play a major role in the polarization transfer.
We set $f_\text{n}$ = 1~MHz (corresponding to 23.4866~mT for protons), $A_\perp$ = 0.3~MHz, $f_\text{t} = f_\text{p}$, and $N_\text{pol}$ = 1.
At the end of PulsePol, the electron spin ends up in either flipped (polarization transferred) or unflipped.
The transition probabilities $\hat{T}_{\alpha}=\bra{-\frac{1}{2}} \hat{U} \ket{+\frac{1}{2}}$ and $\hat{T}_{\beta}=\bra{+\frac{1}{2}} \hat{U} \ket{+\frac{1}{2}}$ of the electron spin can be broken down into
\begin{eqnarray*}
\hat{T}_{\alpha} &=& \alpha_{+} ( \hat{\sigma}_+^{(1)} \mathalpha{+} \hat{\sigma}_+^{(2)} ) \mathalpha{+} \alpha_{-} ( \hat{\sigma}_-^{(1)} \mathalpha{+} \hat{\sigma}_-^{(2)} ) \mathalpha{+} \\
& & \alpha_{-,z} ( \hat{\sigma}_{-}^{(1)}\hat{\sigma}_{z}^{(2)} \mathalpha{+} \hat{\sigma}_{z}^{(1)}\hat{\sigma}_-^{(2)} )
\mathalpha{+} \alpha_{+,z} ( \hat{\sigma}_{+}^{(1)}\hat{\sigma}_{z}^{(2)} \mathalpha{+} \hat{\sigma}_{z}^{(1)}\hat{\sigma}_+^{(2)} ) \\
\hat{T}_{\beta} &=& \beta_\text{e} \bm{1} \mathalpha{+} \beta_{z} (\hat{\sigma}_z^{(1)} \mathalpha{+} \hat{\sigma}_z^{(2)}) \mathalpha{+} \beta_{z,z} \hat{\sigma}_z^{(1)}\hat{\sigma}_z^{(2)} \mathalpha{+} \\
& & \beta_{+,-} (\hat{\sigma}_+^{(1)}\hat{\sigma}_-^{(2)} \mathalpha{+} \hat{\sigma}_-^{(1)}\hat{\sigma}_+^{(2)})
\mathalpha{+} \beta_{+,+} (\hat{\sigma}_+^{(1)}\hat{\sigma}_+^{(2)} \mathalpha{+} \hat{\sigma}_-^{(1)}\hat{\sigma}_-^{(2)}),
\end{eqnarray*}
where $\hat{\sigma}_z = 2\hat{I}_z$ is the Pauli operator and $\hat{\sigma}_{\pm}=(\hat{I}_x \pm i\hat{I}_y)$ are the ladder operators, both on the nuclear spins, and $\bm{1}$ is the identity operator.
The coefficients $\alpha_j$ and $\beta_j$ are the complex transition amplitudes, and the subscript $j$ indicates the corresponding nuclear spin dynamics, as sorted out in Fig.~\ref{fig1}(b).
The explicit forms of the transition amplitudes are given in Supplemental Material~\cite{SM}.

Six dynamics involving the ladder operators [except ``phase flip'' in Fig.~\ref{fig1}(b)] are directly responsible for the (de)polarization,
and their transition amplitudes as a function of $\tau_\text{pol}$, centered at the resonance, are shown in Fig.~\ref{fig1}(c).
$\alpha_{\pm}$ involve only one nuclear spin, with $\abs{\alpha_-}$ taking its maximum and $\abs{\alpha_+}$ falling to zero (node) near the resonance~\cite{Footnote_1}.
This is consistent with the average Hamiltonian treatment examined in previous studies~\cite{SST+18,SIA18}.
Were it not for other dynamics, the nuclear spin could be perfectly polarized in the direction anti-parallel to the magnetic field.

\begin{figure}
\begin{center}
\includegraphics{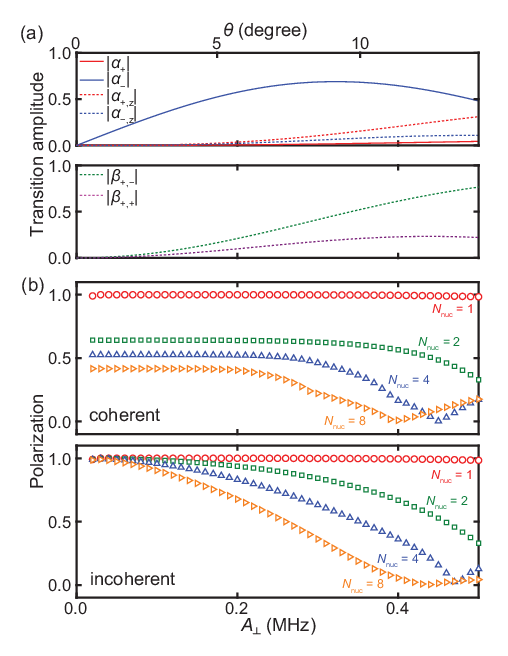}
\caption{
(a) The absolute transition amplitudes of six spin dynamics as a function of $A_\perp$ ($\theta$) at the resonance condition.
(b) The simulated polarizations for the coherent (upper panel) and incoherent (lower panel) cases.
The simulation parameters are the same as in Fig.~\ref{fig1}(c).
\label{fig2}
}\end{center}
\end{figure}

The situation changes when the dynamics involving two spins are in action.
The most critical is the flip-flop that is accompanied by the phase flip of the second nuclear spin ($\alpha_{+,z}$).
This behavior, counteracting the polarization process of $\alpha_-$, is similar to $\alpha_+$, but with non-zero transition amplitude near the resonance.
It is observed from Fig.~\ref{fig1}(c) that the different node positions of $\abs{\alpha_{+,z}}$ and $\abs{\alpha_-}$ make the coexistence of the flip-flop and flip-flip processes unavoidable.
In addition, even when the electron spin is not flipped, there exists a process in which a flip-flip between the two nuclear spins occurs ($nn$ flip-flip, $\beta_{+,+}$).
This process originates from a virtual electron--nuclear flip-flop followed by another virtual electron--nuclear flip-flip, thus bringing the electron spin back to the initial state.
The $nn$ flip-flip suppresses the polarization as it changes the polarization of the nuclear spins regardless of the electron spin state.
For $N_\text{nuc} \geq 2$, all the possible dynamics that arise from the combination of flip-flip and flip-flop can occur.
Figure~\ref{fig2}(a) shows $A_\perp$-dependence (equivalently, $\theta$-dependence) of the transition amplitudes.
All the coefficients grow gradually as increasing $A_\perp$.
$\abs{\alpha_-}$ reaches its maximum at $A_\perp$ = 323~kHz,
where we obtain $\abs{\alpha_-} : \abs{\alpha_{+}} : \abs{\alpha_{+,z}} : \abs{\beta_{+,+}} = 0.69 : 0.02 : 0.13: 0.19$.
Given that $\beta_{+,\pm}$ ($\alpha_+$ and $\alpha_{\pm,z}$) are the second (third) order in $A_\perp$~\cite{SM}, the ratio suggests that depolarization due to higher-order spin dynamics is no longer negligible.

Up to now, we have relied on the analytical expressions for $N_\text{nuc} = 2$ to gain physical insight into the higher-order spin dynamics.
We turn to numerical calculation of the nuclear polarization for $N_\text{nuc} \geq 2$.
The computational procedure involves
(i) the preparation of the initial density matrix $\rho$ with the electron spin in the pure $m_S = + \frac{1}{2}$ state and $N_\text{nuc}$ nuclear spins in a mixed state of $\bm{1}/2^{N_\text{nuc}}$,
(ii) the time evolution $\rho' = \hat{U} \rho \hat{U}^{\dagger}$, and 
(iii) the partial trace over the electron spin state $\rho_\text{n}' = \mathrm{tr}_\text{e}(\rho')$.
After (iii), we evaluate the polarizations of the individual nuclear spins as $|2\langle \hat{I}_z^{(l)} \rangle|$,
and the total polarization $P$ as $|\sum_l 2 \langle \hat{I}_z^{(l)} \rangle/N_\text{nuc}|$.
We rerun (i)--(iii), using $\rho_\text{n}'$ as the new initial state of the nuclear spins.
The procedure is repeated $N_\text{rep} = 1000$ times to reach a saturated value.
We compute $P$ for $N_\text{nuc} = 1, 2, 4, 8$ with uniform $A_\perp$ up to 500~kHz.
Matrix calculations were performed using MATLAB on a 56-core Intel CPU workstation.

It is known that in a coherent nuclear spin ensemble the collective dark state develops and saturates the polarization~\cite{TML03,IKTZ03,TIL03,VCC20,VCP+20}.
This effect is noticeable in the case of the homogeneous hyperfine coupling,
because a flip of one nuclear spin, without the knowledge on which nuclear spin to have flipped, necessarily creates a quantum superposition across the ensemble.
The higher-order spin dynamics and the formation of the dark state can spread cooperatively, and it is not straightforward to clearly separate the two effects.
Nonetheless, as a way to formally distinguish them, we also evaluate $P$ of $\text{diag}(\rho_\text{n}')$,
where $\mathrm{diag}$ is defined as the operation that zeros all the off-diagonal elements of the density matrix.
$\text{diag}(\rho_\text{n}')$ is then fed into the next run.
In this manner, the system is incoherent and no dark state is formed.

Figure~\ref{fig2}(b) shows the simulated polarizations, with the upper (lower) panel for the coherent (incoherent) case.
It is observed that the coherent case exhibits suppressed polarizations even for small $A_\perp$ and $N_\text{nuc}$, which is primarily due to the formation of the dark state.
However, even in the incoherent case, the polarization drops quickly as increasing $A_\perp$.
This is consistent with Fig.~\ref{fig2}(a), in which $\alpha_+$ and $\alpha_{+,z}$ increase with $A_\perp$.
The effect is more pronounced for larger $N_\text{nuc}$.
On the other hand, in both cases, the polarization of $N_\text{nuc} = 1$ is kept at (nearly) unity regardless of $A_\perp$.
This corroborates our assertion that the origin of the observed imperfect polarizations is many-body in nature.

Having analyzed the roles of many-body effects in a model system, we are in a position to examine a more concrete system, the NV center in diamond.
We first perform simulations with realistic conditions, prior to analyzing real experimental data.
An important modification is that the electronic spin of the negatively charged NV center is $S = 1$.
Here, we use the $m_S = -1$ and 0 states, with the latter being free from the hyperfine interaction.
The definition of $\theta$ is changed, accordingly, to $\theta = \arctan(A_\perp/f_\text{n})$.

\begin{figure}
\begin{center}
\includegraphics{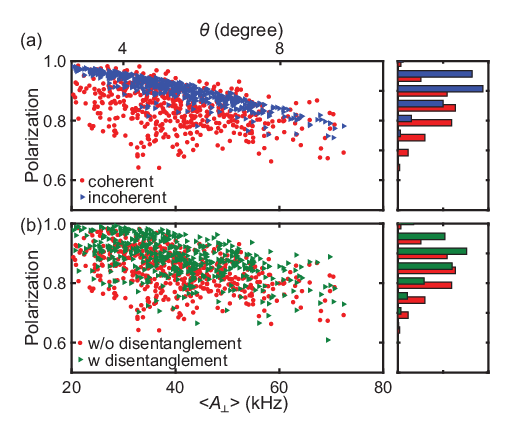}
\caption{
Scatter plots of simulated polarizations of $^{13}$C nuclei with $N_\text{nuc} = 6$.
500 different, randomly generated configurations are shown.
(a) Comparison between the coherent and incoherent cases.
(b) Comparison with and without the disentangling operation.
The right panels show histograms.
\label{fig3}
}\end{center}
\end{figure}

We computationally prepare 500 different configurations of $^{13}$C clusters with $N_\text{nuc}$ = 6 (see~\cite{SM} for detail)
and simulate their polarization dynamics at $B_0$ = 40~mT, a typical value for experiments with NV centers.
Figure~\ref{fig3}(a) shows the scatter plot of simulated polarizations comparing the coherent and incoherent cases.
The horizontal axis $\langle A_\perp \rangle$ is defined as $\sum_{l=1}^{N_\text{nuc}} A_\perp^{(l)}/N_\text{nuc}$ in each cluster.
There is a common tendency that the polarization diminishes with $\langle A_\perp \rangle$,
but the achievable polarizations strongly depend on the details of the hyperfine parameters.
We also observe that the coherent case has a broader distribution than the incoherent case.
The histogram in the right-hand side is peaked at 0.84$\pm$0.07 (0.90$\pm$0.05) in the coherent (incoherent) case.
The low polarization points in the coherent case (for similar $\langle A_\perp \rangle$) presumably arise from the formation of the dark state as in the simulation of Fig.~\ref{fig2}(b),
whereas we attribute the gradual decrease in the polarization for larger $\langle A_\perp \rangle$ to higher-order spin dynamics.
We expect that the effect will become even stronger in larger systems ($N_\text{nuc} >$ 10) of recent experimental interest~\cite{UTW+18,RBvdG+21,BRA+19,ARB+19,CHZ+22}.

In spin systems with disordered hyperfine interactions, the influence of the dark state may be mitigated by introducing a {\it disentangling} operation~\cite{RGG+20}.
A simple realization in the present setup is to reinitialize the NV center into the $m_S = -1$ state (optical initialization followed by a microwave $\pi$ pulse) after PulsePol
and add a wait time of $2 N_\text{pol} \tau_\text{pol}$, during which the respective nuclear spins will precess under different hyperfine fields and acquire different phases.
This modified sequence is tested in Fig.~\ref{fig3}(b) and shows clear improvement from the original sequence, with the peak of the histogram shifted from 0.84$\pm$0.07 to 0.88$\pm$0.07.
The decreasing polarizations with $\langle A_\perp \rangle$, attributable to the higher-order dynamics, are still discernible.

\begin{figure}
\begin{center}
\includegraphics{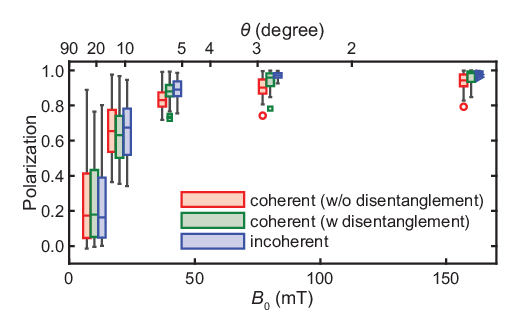}
\caption{
Box chart of $B_0$-dependent polarizations.
Each box consists of 50 elements and is bounded by the upper and lower quartiles, with the median given by the horizontal line.
The minimum and maximum values are indicated by the whiskers, with a few outliers given by circles and triangles.
$N_\text{rep}$ = 500 at 10, 20 and 40~mT, $N_\text{rep}$ = 2000 at 80~mT, and $N_\text{rep}$ = 4000 at 160~mT are chosen to reach saturations.
The plots are shifted horizontally for the purpose of visibility.
The upper horizontal axis is estimated as $\theta = \arctan(\overline{A_\perp}/f_\text{n})$, where $\overline{A_\perp}$ is the average of $\langle A_\perp \rangle$ over 50 configurations.
\label{fig4}
}\end{center}
\end{figure}

It is of practical importance to examine how large the magnetic field should be to obtain near-unity hyperpolarization.
The role of $A_\perp$ should become marginal as increasing $B_0$.
Figure~\ref{fig4} shows $B_0$-dependence of the polarizations for 50 sets of $^{13}$C clusters with $N_\text{nuc}$ = 6.
In the incoherent case, the polarizations at $B_0 = 160$~mT fall on a narrow range of 0.99$\pm$0.01.
On the contrary, the coherent cases show relatively large variations across different clusters, even when the disentangling operation is used.
This reflects, as observed in Fig.~\ref{fig3}, the strong dependence of the achievable polarizations on the details of the hyperfine parameters.
Therefore, in a real system, applying large field of $B_0\gg 40$~mT may not be sufficient; 
a careful selection of $^{13}$C clusters with weak influence of the dark state will be crucial to achieve near-unity hyperpolarization with PulsePol.

The last analysis deals with the experimental data of \cite{RBvdG+21}, 
in which PulsePol was applied to 9 nuclei in a relaxation-free cryogenic environment.
The averaged polarization was 0.64, but the calculation up to $N_\text{nuc}$ = 5 did not reproduce this value~\cite{RBvdG+21}.
In Table~\ref{tab1}, the first two rows list the hyperfine parameters of these 9 nuclei, numbered in descending order of the values of $A_{\perp}$.
\begin{table}
\caption{Experimental hyperfine parameters $A_\perp$ and $A_\parallel$ (in kHz) of 9 spins measured in \cite{RBvdG+21} and simulated polarizations.
coh.: coherent, incoh.: incoherent.}
\label{tab1}
\centering
\begin{tabular}{|l||c|c|c|c|c|c|c|c|c||c|}
\hline
& 1 & 2 & 3 & 4 & 5 & 6 & 7 & 8 & 9 & Ave.\\
\hline
$A_\perp$ & 59.2 & 13.0 & 9.0 & 9.0 & 8.6 & 7.0 & 5.3 & 5.0 & 5.0 & 13.5 \\
\hline
$-A_\parallel$ & 11.3 & 14.1 & 48.6 & 14.0 & 17.6 & 4.7 & 19.8 & 9.8 & 5.6 & 12.2 \\
\hline
\hline
coh. & 0.97 & 0.82 & 0.99 & 0.79 & 0.98 & 0.61 & 0.98 & 0.22 & 0.53 & 0.77 \\
\hline
incoh. & 0.99 & 0.97 & 0.95 & 0.97 & 0.96 & 0.97 & 0.97 & 0.97 & 0.97 & 0.97 \\
\hline
\end{tabular}
\end{table}
We perform a full simulation with 9 spins, using the same parameter set as \cite{RBvdG+21}
($B_0$ = 40.3553~mT, $N_\text{pol}$ =2, $\tau_\text{pol}$ = 8$\times$0.436~$\mu$s, $N_\text{rep}$ = 5000).
The results are shown in the third and fourth rows.
The average polarization in the coherent case is 0.77, close to the result of \cite{RBvdG+21}, with variations across 9 nuclear spins.
High polarizations are obtained in the incoherent bath, consistent with small $\langle A_\perp \rangle$ of 13.5~kHz.
More discussions are given in \cite{SM}.

In summary, we analyzed the impact of higher-order nuclear spin dynamics induced by the pulsed DNP sequence,
in conjunction with the formation of the collective dark state, both many-body in nature, on the nuclear polarization.
The usefulness and limitation of disentangling operations were also discussed. 
We were able to understand the nuclear polarizations observed in the system of a single NV center in diamond reported in \cite{RBvdG+21}.
While we focused on a particular DNP sequence (PulsePol), we show in \cite{SM} that these dynamics are also responsible for another widely-used sequence (NOVEL~\cite{HDSW88}).
In future studies, the effect of finite pulse lengths on higher-order spin dynamics will require a careful analysis.
Systematic Hamiltonian engineering, applicable to multi-spin systems, may be used to achieve higher polarization even at low magnetic fields~\cite{CZK+20}.
We also remark that, via an appropriate control over environmental nuclear spins which otherwise cause decoherence of an electron spin qubit,
they can be converted into a resource valuable to quantum technology~\cite{TML03,IKTZ03,TIL03,RTP+08,CMT+13,GZB+21}.
Our work sheds light on collective engineering of nuclear spin ensemble as well as future designs of DNP protocols.

K.S. was supported by JSPS Grant-in-Aid for Scientific Research (KAKENHI) Grant No.~JP22K03524.


%

\cleardoublepage
\onecolumngrid
\renewcommand{\thefigure}{S\arabic{figure}}
\renewcommand{\theequation}{S\arabic{equation}}
\renewcommand{\thetable}{S\arabic{table}}
\setcounter{figure}{0}
\setcounter{equation}{0}
\setcounter{table}{0}

\part{\uppercase{S}\lowercase{upplemental }\uppercase{M}\lowercase{aterial for ``}\uppercase{S}\lowercase{uppression of }\uppercase{P}\lowercase{ulsed }\uppercase{D}\lowercase{ynamic }\uppercase{N}\lowercase{uclear }\uppercase{P}\lowercase{olarization by }\uppercase{M}\lowercase{any-}\uppercase{B}\lowercase{ody }\uppercase{S}\lowercase{pin }\uppercase{D}\lowercase{ynamics''}}

\section{\uppercase{C}\lowercase{alculation of }\uppercase{NV--$^{13}$C}\lowercase{ hyperfine parameters}}
Naturally-occurring $^{13}$C nuclei ($I = \frac{1}{2}$) occupy 1.1\% of the diamond lattice, whereas the rest is $^{12}$C ($I$ = 0). 
In simulating the DNP process of $^{13}$C nuclei in diamond polarized via the electronic spin of the NV center,
we computationally prepare a diamond lattice with the lattice constant of 0.367~nm (the bond length of 0.155~nm).
The origin is set as the vacancy site of the NV center and the size of the crystal is 2~nm in radius, which contains 5849 carbon sites.
A Monte Carlo method assigns a given site as $^{13}$C with the probability of 1.1\%.

The dipolar interactions between the electronic spin of the NV center and the $^{13}$C nuclear spin are given as
\begin{eqnarray}
A_{\parallel} &=& \frac{\mu_0 h \gamma_{\mathrm{e}}\gamma_{\mathrm{n}} }{4\pi r^3} \left( 3 \cos{^2\theta_r} -1 \right) \\
A_{\perp} &=& \frac{\mu_0 h \gamma_{\mathrm{e}}\gamma_{\mathrm{n}} }{4\pi r^3} \left( 3 \cos{\theta_r}\sin{\theta_r} \right),
\end{eqnarray}
where $r = |\bm{r}|$ is the distance between the vacancy site and the $^{13}$C nucleus, $\theta_r$ is the angle between the two vectors $\bm{r}$ and $\bm{B}_0$,
and $\gamma_\text{e}$ ($\gamma_\text{n}$) is the gyromagnetic ratio of the electronic (nuclear) spin.
For a given configuration, we list up $^{13}$C spins in descending order of the values of $A_\perp$.
When there is a $^{13}$C nucleus very close to the NV center, the hyperfine interaction can be as large as the precession frequency.
In such a case, the resonance condition may not work in the presence of large detuning and the signals of other $^{13}$C spins may be suppressed. 
Therefore, among the randomly generated spin clusters, we use ones that only contain nuclear spin with $A_\perp <$ 100~kHz.

To be precise, the other type of electron--nuclear hyperfine interaction, the Fermi contact interaction, is present, which adds to $A_\parallel$.
However, it does not contribute to $A_\perp$ when $B_0$ is perfectly aligned with the NV symmetry axis (as we assume in our simulations), and thus is not considered. 
The dipole interaction between a pair of nuclei within a cluster is not considered either (see the main text).

$B_0$ is set as 40~mT, a typical value for experiments with NV centers.
Since the gyromagnetic ratio of $^{13}$C is about one-fourth of that of $^{1}$H, $f_\text{n}$ at this magnetic field is 428~kHz.

\section{\uppercase{S}\lowercase{imulation of experimental data in [7]}}
\begin{figure}
\begin{center}
\includegraphics{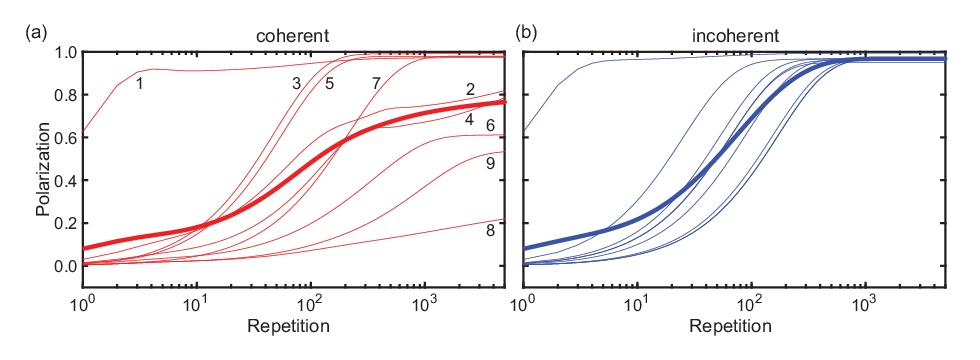}
\caption{Simulations of the polarization buildup of the nine spins in [7] in the coherent (a) and incoherent (b) cases.
The thick line is the average over the nine spins.
The polarization values at $N_\text{rep}$ = 5000 are given in Table~I of the main text.
\label{figs1}
}\end{center}
\end{figure}

In the main text, we showed the values of simulated nuclear polarizations of nine spins reported in [7] (of the main text).
Here, we provide additional data and discussions.
Figure~\ref{figs1} shows the buildup of the nuclear polarizations as a function of $N_\text{rep}$, the number of the repetition of the PulsePol cycle.
In the incoherent case (b), the polarizations of all spins grow up to near-unity after $N_\text{rep} \approx$ 1000.
On the other hand, in the coherent case (a), the polarization buildup of a subset of the spins (\#2, 4, 6, 8, 9) is significantly slow.
Some of them (\#2, 4, 8) do not appear to saturate even at $N_\text{rep}$ = 5000, while others (\#6, 9) saturate but at low polarization values.
This comparison is a clear sign that many-body nuclear spin dynamics are responsible for the suppressed nuclear polarizations.

To gain more insights, we examine the quantum states of a pair of spins within a cluster at $N_\text{rep}$ = 5000.
We calculate the density matrix $\rho_{p,q}$ for spins \#$p$ and \#$q$ ($p \neq q$), defined as a partial trace of the density matrix of the coherent nuclear spin system $\rho'_\text{n}$
(at $N_\text{rep}$ = 5000) over the spins except for $p$ and $q$:
\begin{equation}
\rho_{p,q} = \mathrm{tr}_{\notin{p,q}}(\rho'_\text{n}).
\end{equation}
For instance, the density matrix for spins \#1 and \#2 is given as
\begin{equation}
\rho_{1,2} = 
\begin{pmatrix}
0 & 0 & 0 & 0.02 + 0.01 i \\
0 & 0.01 & -0.01 & -0.03 \\
0 & -0.01 & 0.09 & -0.01 \\
0.02 -0.01 i & -0.03 & -0.01 & 0.90 \\
\end{pmatrix}.
\end{equation}
Denoting the nuclear spin states as $\ket{\uparrow}$ and $\ket{\downarrow}$, we observe that, though $\ket{\downarrow\downarrow}$ is dominant as expected from Fig.~\ref{figs1}(a),
the off-diagonal elements make contributions comparable to the rest of the diagonal elements.
Here, we focus on the two off-diagonal elements $\rho_{p,q}(3,2)$ and $\rho_{p,q}(4,1)$,
which, respectively, are constituents of the anti-parallel nuclear spin states $\ket{\uparrow\downarrow}\pm\ket{\downarrow\uparrow}$ (related to the dark state) and the parallel nuclear spin states $\ket{\uparrow\uparrow}\pm\ket{\downarrow\downarrow}$.

\begin{figure}
\begin{center}
\includegraphics{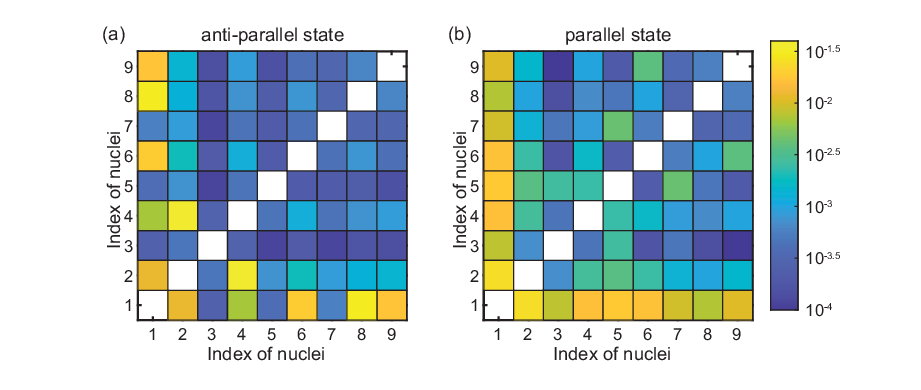}
\caption{
(a) $|\rho_{p,q}(3,2)|$ and (b) $|\rho_{p,q}(4,1)|$ for all combinations of $(p, q)$.
\label{figs2}
}\end{center}
\end{figure}

Figure~\ref{figs2} shows the color plots of $|\rho_{p,q}(3,2)|$ and $|\rho_{p,q}(4,1)|$ for all combinations of $(p,q)$.
It is observed that the off-diagonal elements arise largely from the coupling between spin \#1 and other spins.
Spin \#1 has exceptionally large $A_\perp$ among them [Table~I of the main text] and thus has the largest transition probabilities, both in the first and higher orders.
The first-order dynamics ($\alpha_{-}$) polarize it into $\ket{\downarrow}$ within a few repetitions [Fig.~\ref{figs1}(a)],
but its polarization overshoots at $N_\text{rep} >$ 4 and is recovered only gradually after a number of repetitions.
This behavior is attributable to the higher-order dynamics and the formation of the dark state; the stronger coupling with other spins act to depolarize spin \#1.
As a result, the final polarization of spin \#1 (0.97) is smaller than spins \#3 (0.99), \#5 (0.98), and \#7 (0.98) [Table~I of the main text].
Characteristic of these spins with high polarizations is that they have a comparatively large $A_\parallel$ value (a larger shift in $f_\text{p}$) [Table~I of the main text]
and have a smaller $|\rho_{1,q}(3,2)|$ element [Fig.~\ref{figs2}(a)].
The nuclear spin dynamics in Fig.~\ref{figs1} are qualitatively described as follows.
When the polarized spin \#1 and another still-unpolarized spin ($\ket{\downarrow\uparrow}$) couple via $nn$ flip-flop ($\beta_{+,-}$),
an anti-parallel spin state $\ket{\downarrow\uparrow} + c \ket{\uparrow\downarrow}$, with $c$ the (relative) coefficient determined by $\beta_{+,-}$, may be created.
This process leads to the formation of a dark state, limiting the achievable polarizations.
In the cases of spins \#3, 5, and 7, large $A_\parallel$ results in non-vanishing $\alpha_{\pm,z}$ at the resonance condition ($f_\text{t} = 3/2\tau_\text{pol}$).
Such a phase-flip process can break a dark state ($\ket{\downarrow\uparrow} - \ket{\uparrow\downarrow} \to \ket{\downarrow\uparrow} + \ket{\uparrow\downarrow}$),
allowing for further buildup of the polarizations.
With more than two nuclear spins ($N_\text{nuc} >$ 2), the overall dynamics involve a process in which a pair of anti-parallel spins is swapped with another pair of spins via $nn$ flip-flop.
Still, spins \#3, 5, and 7 are able to escape from being trapped in a dark state via phase-flip.

\section{\uppercase{D}\lowercase{isentangling operation}}
\begin{figure}
\begin{center}
\includegraphics{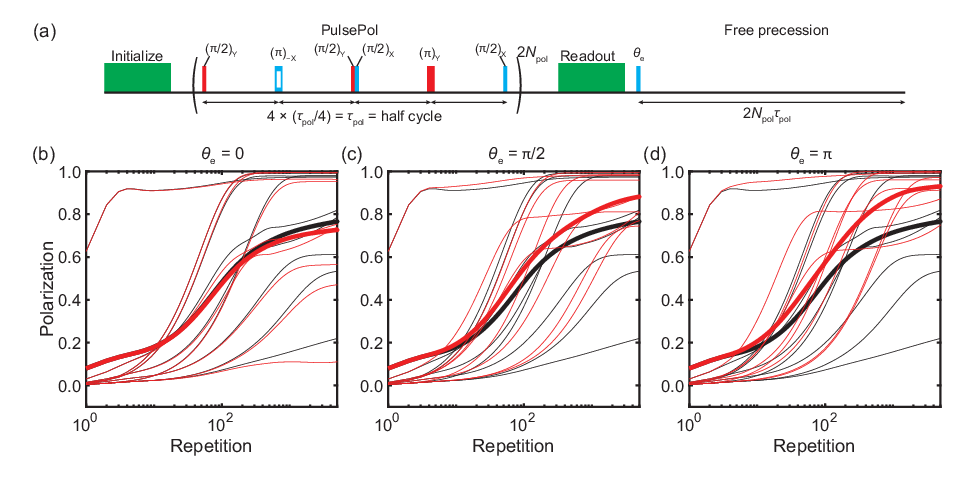}
\caption{
(a) PulsePol with optical initialization/readout (in green) and disentangling operation.
$\theta_\text{e}$ is the rotation angle of the electronic spin of the NV center.
(b--d) Simulations of the polarization buildup of the nine spins in [7] when a disentangling operation with
(b) $\theta_\text{e}$ = 0, (c) $\theta_\text{e} = \pi/2$, and (d) $\theta_\text{e} = \pi$ is applied (red curves).
The original data without disentanglement [Fig.~\ref{figs1}(a)] are also shown for comparison (black curves).
The thick lines are the average over the nine spins. 
The polarizations after $N_\text{rep}$ = 5000 are given in Table~\ref{tabs1}.
\label{figs3}
}\end{center}
\end{figure}

In the main text, we introduced a disentangling operation to partly mitigate the influence of the dark state.
The full sequence is shown in Fig.~\ref{figs3}(a).
$\theta_\text{e}$ is the rotation angle of the electron spin.
$\theta_\text{e} = \pi$ (bringing the NV electronic spin into the $m_S = -1$ state) was adopted in the main text, but in principle the angle can be chosen arbitrarily.
In the case of the NV center, the strength of the hyperfine field is unequal with regard to the two levels $m_S = -1, 0$ and we expect $\theta_\text{e} = \pi$ to be most effective for the phase randomization.

Figures~\ref{figs3}(b--d) show the simulated polarization dynamics of the nuclear spin cluster of [7] under disentanglement.
The polarization values are summarized in Table~\ref{tabs1}.
\begin{table}
\caption{Simulated polarizations after $N_\text{rep}$ = 5000 in Fig.~\ref{figs3}(b--d).}
\label{tabs1}
\centering
\begin{tabular}{|l||c|c|c|c|c|c|c|c|c||c|}
\hline
& 1 & 2 & 3 & 4 & 5 & 6 & 7 & 8 & 9 & Ave.\\
\hline
$\theta_\text{e}$ = 0 & 0.96 & 0.75 & 0.99 & 0.74 & 0.99 & 0.56 & 0.95 & 0.11 & 0.47 & 0.72 \\
\hline
$\theta_\text{e} = \pi/2$ & 0.99 & 0.88 & 0.96 & 0.81 & 0.99 & 0.81 & 0.98 & 0.77 & 0.74 & 0.88 \\
\hline
$\theta_\text{e} = \pi$ & 0.99 & 0.87 & 0.96 & 0.75 & 1.00 & 0.92 & 1.00 & 0.97 & 0.91 & 0.93 \\
\hline
\end{tabular}
\end{table}
As expected, the average polarization over nine spins becomes higher by exerting stronger hyperfine interactions.
In particular, spins \#6, \#8, and \#9, which had small polarizations without the disentangling operation, achieve significantly higher polarizations.
Consequently, the overshoot of polarization in spin \#1, which we interpreted as due to the formation of the dark state with other spins, is weakened for $\theta_\text{e} = \pi/2$ and vanishes for $\theta_\text{e} = \pi$.
Interestingly, the polarizations of spins \#2, \#3, and \#4 are higher when the $\theta_\text{e} = \pi/2$ pulse is used for disentanglement.
Future study to understand the mechanism of this behavior and the role of the rotation angle will help further enhancing the overall nuclear polarization.

\section{\uppercase{A}\lowercase{ case of }\uppercase{NOVEL}}
The spin dynamics of another DNP sequence NOVEL, a widely used protocol in DNP-NMR spectroscopy, are examined for comparison.
The NOVEL Hamiltonian is defined as 
\begin{equation}
\hat{H}_\text{NOVEL} = f_\text{t} \hat{S}_y + \sum_{l = 1}^{N_\text{nuc}} - f_\text{n} \hat{I}_z^{(l)} + A_\perp^{(l)} \hat{S}_z \hat{I}_x^{(l)},
\end{equation}
where $f_\text{t}$ is the target frequency and $\hat{S}_{y,z}$ are the $S = \frac{1}{2}$ electron spin operators.
The unitary evolution is given by
\begin{equation}
\hat{U}_\text{NOVEL} = \hat{U}_x e^{ -2\pi i \hat{H}_\text{NOVEL} t_\text{dur}} \hat{U}_x^\dagger,
\end{equation}
where $ t_\text{dur}$ is the duration of the sequence.
To compare with PulsePol, the parameters are set as $f_\text{t} = 3/2 \tau_\text{pol}$ and $t_\text{dur} = N_\text{pol} \times 2 \tau_\text{pol}$, with $\tau_\text{pol}$ and $N_\text{pol}$ as used in PulsePol.

\begin{figure}
\begin{center}
\includegraphics{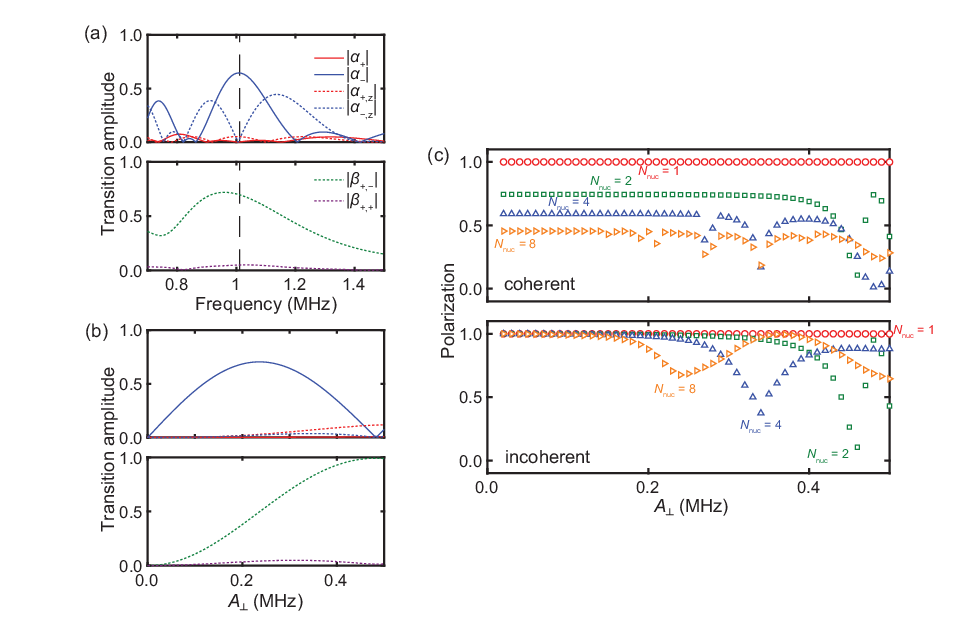}
\caption{
(a) Frequency dependence of the absolute transition amplitudes under NOVEL.
$f_\text{n}$ = 1~MHz and $A_\perp$ = 300~kHz.
The dashed vertical line indicates the precession frequency $f_\text{p} = \sqrt{f_\text{n}^2 + (A_\perp/2)^2}$.
(b) $A_\perp$ dependence of the absolute transition amplitudes at $f_\text{p}$.
(c) The simulated polarizations under NOVEL for the coherent and incoherent cases.
\label{figs4}
}\end{center}
\end{figure}

Figures~\ref{figs4}(a) and (b) show the frequency- and $A_\perp$-dependences of the absolute transition amplitudes, corresponding to Fig.~1(c) and Fig.~2(a) in the case of PulsePol, respectively.
From Fig.~\ref{figs4}(a), we observe that $|\alpha_+|$, $|\alpha_{+,z}|$, and $|\beta_{+,+}|$, the processes that counteract $|\alpha_-|$, also appear in NOVEL,
demonstrating that the higher-order depolarization dynamics we discussed in the main text are not limited to PulsePol.
In NOVEL, $|\alpha_-|$ ($en$ flip-flip) increases steeply with increasing $A_\perp$, and takes its maximum at a smaller value compared with PulsePol,
whereas $|\alpha_+|$, $|\alpha_{+,z}|$, and $|\beta_{+,+}|$ are kept relatively small.
This means that NOVEL is less affected by higher-order spin dynamics at the cost of the operational bandwidth.
For spin systems that do not require the broadband property of PulsePol, NOVEL may be preferred.
Designing a broadband DNP sequence with less influence of higher-order spin dynamics will be an interesting and important future direction.

In Fig.~\ref{figs4}(b), $|\alpha_-|$ increases until $A_\perp \approx$ 0.24~MHz, drops to zero and increases again at $A_\perp \approx$ 0.48~MHz.
Notably, $|\alpha_{+,z}|$ continues to grow with $A_\perp$ and thus at $A_\perp \approx$ 0.48~MHz the depolarization ($en$ flip-flop) overcomes the polarization ($en$ flip-flip),
leading the dip in Fig.~\ref{figs4}(c) for $N_\text{nuc}$ = 2.
This behavior has also been observed in PulsePol, but is more pronounced in NOVEL, especially for $N_\text{nuc} >$ 2, where multiple dips are discernible.
From numerical calculations, we confirmed that the node of $|\alpha_{-}|$ shifts to smaller $A_\perp$ as increasing $N_\text{nuc}$, consistent with the behaviors in Fig.~\ref{figs4}(c).
This result certifies that many-body dynamics play an essential role in NOVEL with multiple nuclear spins.


\section{\uppercase{T}\lowercase{ransition amplitudes}}
Below, we list the explicit forms of the transition amplitudes.
The transition amplitudes when the electron spin is flipped are given by
\begin{eqnarray}
\alpha_+ &=& \alpha_x - i\alpha_y \\
\alpha_- &=& \alpha_x + i\alpha_y \\
\alpha_{+,z} &=& \alpha_{x,z} - i\alpha_{y,z} \\
\alpha_{-,z} &=& \alpha_{x,z} + i\alpha_{y,z} \\
\alpha_x &=& \frac{-1+i}{4096} \notag \\
& & \times \{ 8\cos{(\phi/2)}\sin{^5(\phi/2)} [ -((2958+5514\cos{(\phi)}+4520\cos{(2\phi)} \notag \\
& & +3325\cos{(3\phi)}+1610\cos{(4\phi)}+505\cos{(5\phi)})\sin{(5\theta)}) \notag \\
& & +8(889+1596\cos{(\phi)}+1096\cos{(2\phi)}+532\cos{(3\phi)}+175\cos{(4\phi)})\sin{^2(\phi/2)}\sin{(7\theta)} \notag \\
& & -64\cos{(\phi)}(153+222\cos{(\phi)}+73\cos{(2\phi)})\sin{^4(\phi/2)}\sin{(9\theta)} \notag \\
& & +64(67+96\cos{(\phi)}+33\cos{(2\phi)})\sin{^6(\phi/2)}\sin{(11\theta)} ] \notag \\
& & +\sin{^3(\phi)} [ 4(302-398\cos{(\phi)}+464\cos{(2\phi)}-29\cos{(3\phi)}+130\cos{(4\phi)}+43\cos{(5\phi)})\sin{(\theta)} \notag \\
& & +(598-1258\cos{(\phi)}+1592\cos{(2\phi)}+47\cos{(3\phi)} \notag \\
& & +626\cos{(4\phi)}+443\cos{(5\phi)})\sin{(3\theta)}-1536\sin{^{10}(\phi/2)}\sin{(13\theta)} ] \} \\
\alpha_y &=& \frac{-1-i}{2048}\sin{^4(\phi/2)} \notag \\
& & \times \{ \cos{^2(\phi/2)}[(-278+2042\cos{(\phi)}+2984\cos{(2\phi)} \notag \\
& & +6001\cos{(3\phi)}+3182\cos{(4\phi)}+2453\cos{(5\phi)})\sin{(2\theta)} \notag \\
& & +(-1542+1646\cos{(\phi)}-856\cos{(2\phi)}+4099\cos{(3\phi)}+1886\cos{(4\phi)}+2959\cos{(5\phi)})\sin{(4\theta)} \notag \\
& & -64(115+156\cos{(\phi)}+65\cos{(2\phi)})\sin{^6(\phi/2)}\sin{(10\theta)}
-768(3+7\cos{(\phi)})\sin{^8(\phi/2)}\sin{(12\theta)} ] \notag \\
& & +\sin{^2(\phi)} [(-2775-4168\cos{(\phi)}-4812\cos{(2\phi)}-2808\cos{(3\phi)}-1821\cos{(4\phi)})\sin{(6\theta)} \notag \\
& & +896\sin{^8(\phi/2)}\sin{(14\theta)} ] \notag \\
& & +4\sin{^4(\phi/2)} [ (4331+7496\cos{(\phi)}+5116\cos{(2\phi)} \notag \\
& & +2424\cos{(3\phi)}+601\cos{(4\phi)})\sin{(8\theta)}-64\sin{^8(\phi/2)}\sin{(16\theta)} ] \} \\
\alpha_{x,z} &=& (2-2i)\cos{(\phi/2)}\cos{^4(\theta)}\sin{^5(\phi/2)} [ 18+61\cos{(\phi)}+30\cos{(2\phi)} \notag \\
& & +19\cos{(3\phi)}-4(1+3\cos{(\phi)})(17+11\cos{(\phi)})\cos{(2\theta)}\sin{^2(\phi/2)} \notag \\
& & -16(5+3\cos{(\phi)})\cos{(4\theta)}\sin{^4(\phi/2)} \notag \\
& & +16\cos{(6\theta)}\sin{^6(\phi/2)} ]\sin{(\theta)}[\cos{(\phi)}\cos{^2(\theta)}+\sin{^2(\theta)}]^2 \\
\alpha_{y,z} &=& -\frac{1}{512}i\cos{^2(\phi/2)}\sin{^4(\phi/2)} \notag \\
& & \times \{ (1-i)(1+3\cos{(\phi)})(823+1408\cos{(\phi)}+956\cos{(2\phi)}+704\cos{(3\phi)}+205\cos{(4\phi)})\sin{(2\theta)} \notag \\
& & +(4-4i) [ (1+3\cos{(\phi)})(25+142\cos{(\phi)}+120\cos{(2\phi)}+146\cos{(3\phi)}+79\cos{(4\phi)})\sin{(4\theta)} \notag \\
& & -(1+3\cos{(\phi)})(326+659\cos{(\phi)}+522\cos{(2\phi)}+285\cos{(3\phi)})\sin{^2(\phi/2)}\sin{(6\theta)} \notag \\
& & +32(64+121\cos{(\phi)}+72\cos{(2\phi)}+31\cos{(3\phi)})\sin{^4(\phi/2)}\sin{(8\theta)} \notag \\
& & -8(233+356\cos{(\phi)}+179\cos{(2\phi)})\sin{^6(\phi/2)}\sin{(10\theta)} \notag \\
& & +384(2+3\cos{(\phi)})\sin{^8(\phi/2)}\sin{(12\theta)}-192\sin{^{10}(\phi/2)}\sin{(14\theta)} ] \},
\end{eqnarray}
where $\phi = f_\text{p} t_\text{pol}/4$ is the precession phase and $\theta = \arctan(A_\perp/2f_\text{n})$ is the tilt angle.
Keeping up to the third-order term of $\theta$, we obtain
\begin{eqnarray}
\alpha_+ &\approx& (-4+4i)\theta\cos{(\phi/2)}\sin{(\phi/2)}^3(\cos{(2\phi)}+2\cos{(4\phi)}+\cos{(6\phi)}-\sin{(2\phi)}+\sin{(6\phi)}) \notag \\
& & +\frac{1-i}{3}\theta^3\sin{(\phi/2)}^3 [19\cos{(3\phi/2)}-53\cos{(5\phi/2)}+38\cos{(7\phi/2)}+38\cos{(9\phi/2)} \notag \\
& & -29\cos{(11\phi/2)}+43\cos{(13\phi/2)}+20\sin{(3\phi/2)}-64\sin{(5\phi/2)} \notag \\
& & +60\sin{(7\phi/2)}-60\sin{(9\phi/2)}+16\sin{(11\phi/2)}+28\sin{(13\phi/2)}] \\
\alpha_- &\approx& (-4+4i)\theta\cos{(\phi/2)}\sin{(\phi/2)}^3(\cos{(2\phi)}+2\cos{(4\phi)}+\cos{(6\phi)}+\sin{(2\phi)}-\sin{(6\phi)}) \notag \\
& & +\frac{1-i}{3}\theta^3\sin{(\phi/2)}^3 [ 19\cos{(3\phi/2)}-53\cos{(5\phi/2)}+38\cos{(7\phi/2)}+38\cos{(9\phi/2)} \notag \\
& & -29\cos{(11\phi/2)}+43\cos{(13\phi/2)} -20\sin{(3\phi/2)}+64\sin{(5\phi/2)} \notag \\
& & -60\sin{(7\phi/2)}+60\sin{(9\phi/2)}-16\sin{(11\phi/2)}-28\sin{(13\phi/2)} ] \\
\alpha_{+,z} &\approx& -32(1+i)\theta\cos{(\phi/2)}^2\sin{(\phi/2)}^4(\cos{(\phi)}-\sin{(\phi)})(1+\cos{(2\phi)}+\sin{(2\phi)})^2 \notag \\
& & +\frac{8(1+i)}{3}\theta^3\cos{(\phi/2)}\cos{(\phi)}\sin{(\phi/2)}^4 [8\cos{(\phi/2)}+8\cos{(3\phi/2)}+56\cos{(5\phi/2)} \notag \\
& & -38\cos{(7\phi/2)}+46\cos{(9\phi/2)}+36\sin{(3\phi/2)}-60\sin{(5\phi/2)}+25\left(\sin{(7\phi/2)}+\sin{(9\phi/2)}\right) ] \\
\alpha_{-,z} &\approx& -32(1+i)\theta\cos{(\phi/2)}^2\sin{(\phi/2)}^4(\cos{(\phi)}+\sin{(\phi)})(1+\cos{(2\phi)}-\sin{(2\phi)})^2 \notag \\
& & +\frac{8(1+i)}{3}\theta^3\cos{(\phi/2)}\cos{(\phi)}\sin{(\phi/2)}^4 [ 8\cos{(\phi/2)}+8\cos{(3\phi/2)}+56\cos{(5\phi/2)} \notag \\
& & -38\cos{(7\phi/2)}+46\cos{(9\phi/2)}-36\sin{(3\phi/2)}+60\sin{(5\phi/2)}-25\left(\sin{(7\phi/2)}+\sin{(9\phi/2)}\right) ].
\end{eqnarray}

The transition amplitudes when the electron spin is not flipped are given by
\begin{eqnarray}
\beta_{+,-} &=& \beta_{x,x} + \beta_{y,y} \\
\beta_{+,+} &=& \beta_{x,x} - \beta_{y,y} \\
\beta_{x,x} &=& \frac{1-i}{256}\cos{^2(\phi/2)}\sin{^6(\phi/2)} \notag \\
& & \times \{ 4[(481+128i)+744\cos{(\phi)}+436\cos{(2\phi)}+216\cos{(3\phi)}+43\cos{(4\phi)}] \notag \\
& & +[(653+256i)+952\cos{(\phi)}+1316\cos{(2\phi)}+648\cos{(3\phi)}+271\cos{(4\phi)}]\cos{(2\theta)} \notag \\
& & -2[(931+256i)+1784\cos{(\phi)}+700\cos{(2\phi)}+456\cos{(3\phi)}-31\cos{(4\phi)}]\cos{(4\theta)} \notag \\
& & -[(497+256i)+1176\cos{(\phi)}+1076\cos{(2\phi)}+936\cos{(3\phi)}+155\cos{(4\phi)}]\cos{(6\theta)} \notag \\
& & +16(114+221\cos{(\phi)}+126\cos{(2\phi)}+51\cos{(3\phi)})\cos{(8\theta)}\sin{^2(\phi/2)} \notag \\
& & -16(115+188\cos{(\phi)}+113\cos{(2\phi)})\cos{(10\theta)}\sin{^4(\phi/2)} \notag \\
& & +384(3+5\cos{(\phi)})\cos{(12\theta)}\sin{^6(\phi/2)}-384\cos{(14\theta)}\sin{^8(\phi/2)} \} \\
\beta_{y,y} &=& \frac{1+i}{256}\sin{^4(\phi/2)} [ (167-128i)-24\cos{(\phi)}-340\cos{(2\phi)}-40\cos{(3\phi)}+109\cos{(4\phi)} \notag \\
& & +256\cos{^2(\phi/2)}[(30+34\cos{(\phi)})\cos{(2\theta)} +(9+5\cos{(\phi)})\cos{(4\theta)}]\sin{^4(\phi/2)} \notag \\
& & +1024\cos{^2(\phi/2)}\cos{(6\theta)}\sin{^6(\phi/2)}-128\cos{(8\theta)}\sin{^8(\phi/2)} ] \notag \\
& & \times [-2\cos{^2(\phi/2)}\sin{(2\theta)}+\sin{^2(\phi/2)}\sin{(4\theta)}]^2.
\end{eqnarray}
Keeping up to the third-order term of $\theta$, we obtain
\begin{eqnarray}
\beta_e &\approx& -\cos{(4\phi)}^2+2i\theta^2((1+6i)\cos{(\phi)}-(2-4i)\cos{(2\phi)}+(1+2i)\cos{(3\phi)} \notag \\
& & +8i\cos{(4\phi)}+(1+2i)\cos{(5\phi)}-(2-4i)\cos{(6\phi)}+(1+6i)\cos{(7\phi)})\sin{(\phi/2)}^2 \\
\beta_z &\approx& -i\sin{(8\phi)}+\theta^2 [ 2\sin{(\phi)}-(3+i)\sin{(2\phi)}+(2+4i)\sin{(3\phi)} \notag \\
& & +(1-6i)\sin{(4\phi)}-(6-4i)\sin{(5\phi)}+(9+i)\sin{(6\phi)}-(6+8i)\sin{(7\phi)}+\left(\frac{3}{2}+6i\right)\sin{(8\phi)} ] \\
\beta_{z,z} &\approx& -i\theta^2[(4+4i)\cos{(\phi)}-(6+i)\cos{(2\phi)}+4\cos{(3\phi)}-4\cos{(5\phi)} \notag \\
& & +(6+i)\cos{(6\phi)}-(4+4i)\cos{(7\phi)}+(1+3i)(-1+\cos{(8\phi)})]+\sin{(4\phi)}^2 \\
\beta_{+,-} &\approx& 8\theta^2(2-3i\cos{(2\phi)}-2\cos{(4\phi)}+i\cos{(6\phi)})\sin{(\phi/2)}^4 \\
\beta_{+,+} &\approx& 8\theta^2(\cos{(2\phi)}+2i\cos{(4\phi)}-\cos{(6\phi)})\sin{(\phi/2)}^4.
\end{eqnarray}
At the resonance condition $\phi = 3\pi/4$, $\alpha_\pm$, $\alpha_{\pm,z}$, $\beta_\text{e}$, $\beta_{z}$, $\beta_{z,z}$, $\beta_{+,\mp}$ are simplified as
\begin{eqnarray}
\alpha_+ &\approx& \frac{-1+i}{2} (5+3\sqrt{2})\theta^3 \\
\alpha_- &\approx& 2(1-i)(1+\sqrt{2})\theta - \frac{1-i}{6}(239+173)\theta^3 \\
\alpha_{+,z} &\approx& -2(1+i)(7+5\sqrt{2})\theta^3 \\
\alpha_{-,z} &\approx& -2(1+i)(5+3\sqrt{2})\theta^3 \\
\beta_{e} &\approx& -1 + (12+8\sqrt{2}) \theta^2 \\
\beta_{z} &\approx& [(6+i)+(4+2i)\sqrt{2}] \theta^2 \\
\beta_{z,z} &\approx& 0 \\
\beta_{+,-} &\approx& (12+8\sqrt{2}) \theta^2 \\
\beta_{+,+} &\approx& -2i(3+2\sqrt{2}) \theta^2.
\end{eqnarray}


\begin{thebibliography}{51}
\bibitem [{Foo()}]{Footnote_1}%
  \BibitemOpen
  \href@noop {} {}\bibinfo {note} {In this sequence, the electron--nuclear
  flip-flip is dominant. Instead, the electron--nuclear flip-flop can be made
  dominant by interchanging the phases of the $\pi$/2
  pulses~\cite{SIA18}.}\BibitemShut {Stop}%
\end{thebibliography}
\end{document}